\documentclass[3p, times, twocolumn, sort&compress]{elsarticle}
\usepackage{color}
\usepackage{amssymb}
\usepackage[german,english]{babel}
\usepackage{amsfonts}
\usepackage{amsmath}
\usepackage{array}
\usepackage{epsf}
\usepackage{epsfig}
\usepackage{float}
\usepackage[version=3]{mhchem}
\usepackage{textcomp}
\usepackage{graphics}
\usepackage{graphicx}
\usepackage{epstopdf}
\usepackage{placeins}
\usepackage{footnote}
\usepackage[para]{threeparttablex}
\makesavenoteenv{tabular}
\makesavenoteenv{table}
\newcolumntype{x}[1]{>{\centering\arraybackslash\hspace{0pt}}p{#1}}

\begin{document}
\def\cm{cm$^{-1}$}
\newcommand*{\eref}[1]{Eq.~(\plainref{#1})}
\newcommand*{\fref}[1]{Fig.~\plainref{#1}}
\newcommand*{\tref}[1]{Tab.~\plainref{#1}}
\newcommand{\onlinecite}[1]{\hspace{-1 ex} \nocite{#1}\citenum{#1}} 
\newcommand*{\bref}[1]{Ref.~\onlinecite{#1}}

\begin{frontmatter}
\title{The light-harvesting complex 2 of \textit{Allochromatium  vinosum}: B800 absorption band splitting and exciton relaxation}
%
\author{Xiaomeng Liu}
\author{Oliver K\"uhn}
\ead{oliver.kuehn@uni-rostock.de}
\cortext[cor1]{Corresponding author}
\address
{Institut f\"ur Physik, Universit\"at Rostock, Albert-Einstein-Str. 23-24, D-18059 Rostock, Germany}

\begin{abstract}
\textit{Allochromatium (Alc.) vinosum} has a double-peak structure of its absorption band around 800~nm. Previously, the excitonic origin of this feature has been demonstrated experimentally, but a detailed understanding still lacks a  model Hamiltonian being able to reproduce absorption as well as exciton relaxation time scales. Here, we propose a system-bath model which accommodates  these observables. It combines Frenkel exciton theory for a dimerized and energetically heterogeneous B800 pigment pool with a quantum master equation approach describing phase and energy relaxation according to an experimental spectral density. The analysis of this model shows that the LH2 of \textit{Alc. vinosum} features an interesting interplay of two excitonic bands, which are originating from the different pigment pools. 
\end{abstract}
\end{frontmatter}
\cleardoublepage
\newpage

\section{Introduction}
The light-harvesting antenna complexes of purple non-sulphur photosynthetic bacteria provide prime examples for the importance of quantum effects for biological function~\cite{cogdell06_227,scholes17_647}. Fascination especially among non-biologists has been triggered by the publication of the high-resolution structure for the peripheral antenna LH2 of \textit{Rhodopseudomonas (Rps.) acidophila} by McDermott et al. in 1995~\cite{mcdermott95_517}. The modular design of rings, comprised of nine pairs of $\alpha\beta$-apoproteins, each pair binding three BChl $a$ molecules, facilitates a wealth of scenarios as far as exciton dynamics is concerned (for reviews, see e.g. Refs. \cite{pullerits96_381, kuhn97_213, valkunas00, renger01_137}). 
From the dynamics and spectroscopy point of view the LH2 consists of two different pigment pools, i.e.\  the strongly coupled 18 BChl $a$ molecules forming the B850 ring,  whose bacteriochlorin planes are perpendicular to the transmembrane $\alpha$ helices, and the more weakly coupled 9 BChl $a$ molecules, whose bacteriochlorin planes are essentially perpendicular to the B850 ones. The two pigment pools gives rise to two absorption features at about 800 and 850~nm as indicated by the labeling. It is commonly assumed that the electronic excitation of the B850 pool is rather delocalized and the transfer is of exciton relaxation type~\cite{kuhn97_4154,chachisvilis97_7275}, whereas the B800 pool is characterized by hopping like incoherent transfer~\cite{pullerits97_10560}, although the latter view has been challenged by recent simulations~\cite{smyth15_30805,shibl17_184001}. Different proposals also exist for the inter-pool B800 to B850 transfer. Due to the weak coupling, modified F\"orster theory taking into account the excitonic delocalization seems to be appropriate~\cite{scholes03_57,sener11_518}. However, due to the overlap between the B800 and B850 band states inter-pool coherences could be operative to facilitate the rapid B800-B850 transfer ~\cite{wu96_12022,pullerits97_10560,kuhn97_3432,renger01_137,smyth15_30805,shibl17_184001}. This view has been supported by recent investigations using two-dimensional spectroscopy~\cite{karki18_,tiwari18_4219,schroter18_114107}.

In terms of the absorption spectrum, \textit{Rps. acidophila} is rather typical for purple bacteria. Other commonly studied natural variants feature band shifts or suppression of one band~\cite{cogdell06_227}.  An interesting exception in this respect is \textit{Alc. vinosum}, whose B800 absorption band has a double-peak structure~\cite{kereiche08_3650}. There are two hypotheses concerning the origin of this band splitting into a blue (B800b) and red (B800r) component. First, the two peaks could be due to two structurally slightly different LH2 complexes, similar to what has been found for \textit{Chromatium tepidum}~\cite{vandijk98_1269}. 
A second hypothesis builds on the observation that there are two main $\alpha$-apoprotein types, suggesting that there could be a structural motif with alternating protein subunit types within a single LH2~\cite{carey14_1849}. This could lead to an excitonic dimerization of the B800 pool, i.e. due to  alternating intermolecular distances and/or different monomeric transition energies. L\"ohner et al. have proposed an excitonic model with alternating distances but equal transition energies to simulate their circular dichroism and   polarization-resolved single-molecule spectroscopy data taken at 1.2~K~\cite{lohner15_23}. However, in earlier transient absorption experiments the excitonic coupling  of the B800 bands upon selective excitation of one sub-band was not observed as a simultaneous bleaching signal~\cite{niedzwiedzki12_1576}. Parallel to these findings, hole-burning experiments have been interpreted in terms of weakly and strongly hydrogen-bonded  pigments giving rise to the two B800 sub-bands, including conformational changes due to proton transfer upon illumination~\cite{kell17_4435}. Dimerization has also been invoked in the transient absorption study reported in Ref.~\cite{luer15_1885}.

In a recent investigation of the dynamics of \textit{Alc. vinosum} at 77~K using two-dimensional electronic spectroscopy, Schr\"oter et al.~\cite{schroter18_1340} provided unambiguous evidence for the excitonic coupling between the two sub-bands. Analysis of diagonal and cross-peak evolution time scales for the inter-band transfers have been established as follows: 3.9~ps for B800b$\rightarrow$B800r, 1.0~ps for B800b$\rightarrow$B850, and 1.4~ps for B800r$\rightarrow$B850. Note that these time scales are rather similar to the ones previously reported in Ref.~\cite{luer15_1885} although the B800 double-peak structure has not been very pronounced in the absorption of the studied room temperature case.

The analysis in Ref.~\cite{schroter18_1340} has been based on a global kinetic modeling of an effective three-state system. Thus, although the excitonic nature of the double-peak has been demonstrated, no information could be obtained about the underlying exciton Hamiltonian. This provides the motivation for the present study, which proposes a simple yet non-trivial system-bath model capable of reproducing the linear absorption spectrum and the inter-band exciton population relaxation times for \textit{Alc. vinosum} at 77~K. 

The paper is organized as follows: In Section \ref{sec:methods} we first outline the spatial arrangement of B800 and B850 molecules, thereby following earlier work by L\"ohner at al.~\cite{lohner15_23}. Next, the system-bath approach is briefly introduced, which leads to the identification of four different models to be investigated in Section~\ref{sec:results}. Results are presented for the absorption spectra of the four models as well as for the population dynamics of that model which best fits the experimental results. A summary is provided in Section \ref{sec:summary}.

\section{Theoretical Methods}
\label{sec:methods}
\subsection{Model Systems}

\begin{figure}[t]
\begin{center}
\includegraphics[width=1.1\columnwidth]{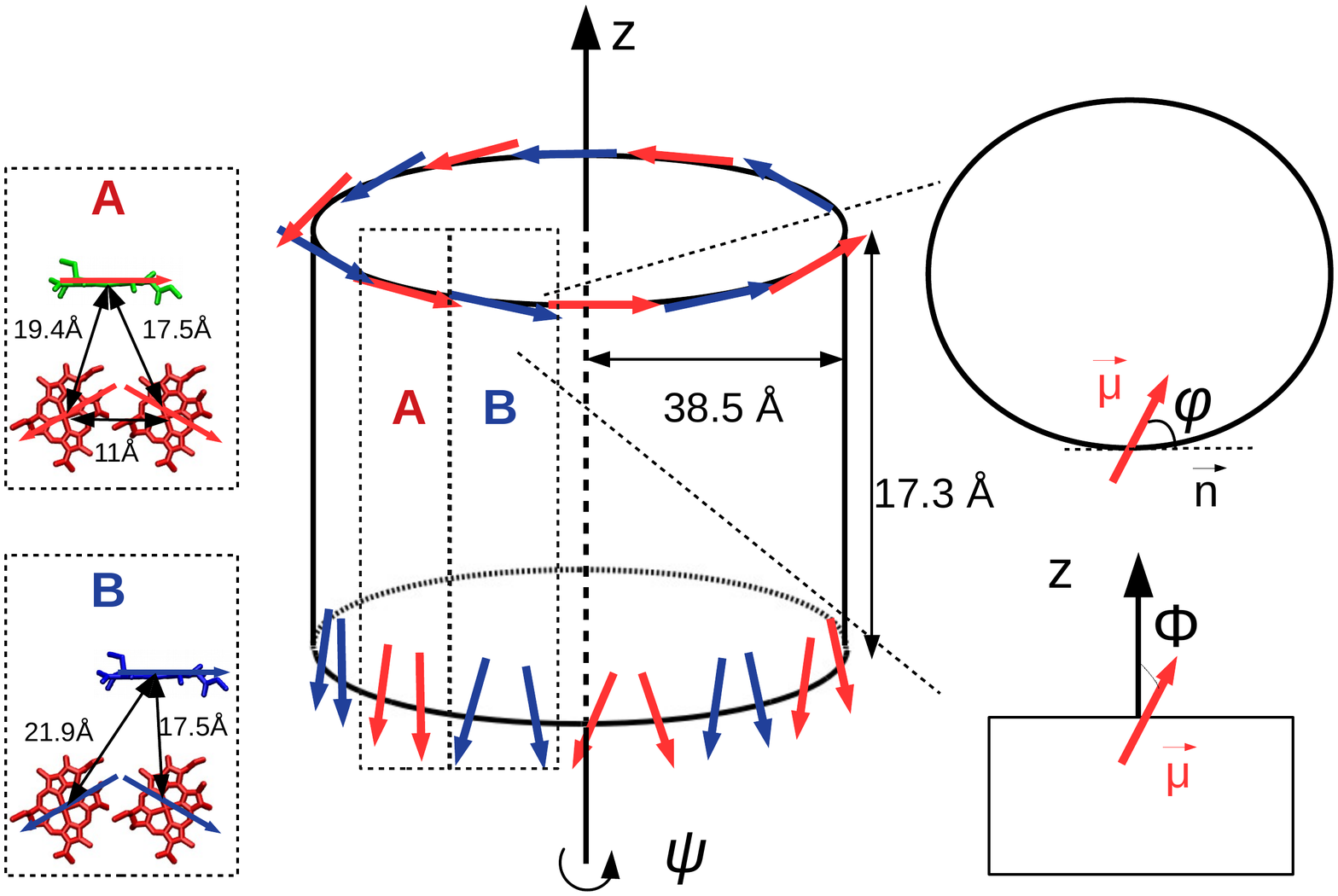}
\caption{ Scheme of the geometry and set of angles for  {BChl} $a$ molecules (B850 (lower) or B800 (upper)) in the LH2 model of L\"ohner et al. \cite{lohner15_23} (for the sake of presentation not all BChl $a$ molecules are shown). The projection of dipole moment $\vec{\mu}$ onto the ring plane and the tangent $\vec{n}$ of the ring makes the angle  $\varphi$. The angle $\phi$ is between $\vec{\mu}$ and direction of the axis $\vec{z}$. Overall, there is a torsional angle $\psi$ between the two rings. In the left part  the two structural motifs, with B800A and B800B molecules, are shown.} 
\label{fig:geometry}
\end{center}
\end{figure}

In Ref.~\cite{lohner15_23} a model starting with a 12-fold symmetry for the arrangement of the B800 and B850 chromophores and in particular for  the direction of the transition dipole moments had been developed using \textit{Rhodospirillum molischianum}~\cite{koepke96_581} as a template. In this model the  36 {BChl} $a$ molecules are arranged in two rings (radius 38.5~\AA{}) as shown in Fig.~\ref{fig:geometry}.  The center to center distance between the B800 (upper) and B850 (lower) ring is 17.3~\AA. The transition dipole moments of  the {BChl} $a$ molecules, $\vec{\mu}_m$, are characterized by two angles: $\varphi_m$ is the angle between the projection of the  dipole moment  and the local tangent $\vec{n}_m$ of the $m$th molecule in the plane of the ring. $\phi_m$ is the angle between the dipole moment and the direction of the cylinder axis $\vec{z}$. Further the two rings are rotated with respect to each other by an angle $\psi$.  Two motifs for the basic B800-B850 units are used, called B800A and B800B; they differ in the position of B800 with respect to B850 as shown in Fig.~\ref{fig:geometry}. (Note that in the following we will use the labels B800A and B800B to distinguish the two type of B800 molecules.) \textcolor{black}{This yields a dimerization of the B800 pool with intra-dimer distances of 15.4~\AA{} and inter-dimer distances of 24.4~\AA. The distance between two  B850 molecules within a motif is 9~\AA{} and between the B800A and B800B motifs it is 11~\AA. } Such a dimerization is  in accord with the observation of two main $\alpha$-apoprotein types with equal abundance~\cite{carey14_1849}.

The fitting of the set of angles by L{\"o}hner et al.~\cite{ lohner15_23} has been performed using the fluorescence excitation spectrum for the complex embedded into a polymer matrix at 1.2 K. The following values have been obtained: $\varphi_{800}=0^{\circ}$ and $\phi_{800}=90^{\circ}$ for B800, $\varphi_{850}=10^{\circ}/170^{\circ}$ and $\phi_{850}=110^{\circ}/70^{\circ}$ for the two B850 molecules within one unit, and $\psi=10^{\circ}$. Based on these geometries, the Coulomb interaction between excitations at different sites has been calculated in dipole approximation. For the monomeric dipoles values of 8.25~D (B800) and 7.5~D (B850) have been used. Further, it was assumed that all site energies for B850 and B800 are equal  to $E_m=$12900 \cm. As far as the line broadening is concerned a simple model of constant linewidths for B800 and B850 was taken. These parameters define \textbf{model 1} of the present study; its excitonic parameters are summarized in Tab.~\ref{tab:lh2parameters}. Notice that due to the tight packing of the overall LH2 structure the maximum couplings are rather large as compared with other LH2 systems. In principle this also questions the validity of the dipole approximation, an issue which will not be further addressed here for simplicity.

\begin{table*}[t]
\caption{Parameters of the different models used in this work. \textcolor{black}{Notice that only the maximum values of the Coulomb interactions are given. The complete Hamiltonian matrices are provided in the Supplementary Material.}}\label{tab:lh2parameters}
\centering
\begin{tabular}{l|c|c|c|c}
 \hline
  & {model 1} & {model 2} & {model 3} &{model 4} \\ \hline
 B800 dipole moment (D)& 8.25 & 8.25 & 8.25 & 8.25   \\
 max. B800-B800 interaction (\cm) & 186 & 186 & 186 & 186   \\
 $\hat{\Gamma}_{800}$ (\cm) & 0 & 0 & 300 & 300   \\
$\Delta E_{\rm 800B}$  (\cm) & 0 & -250 & -250 &-250 \\
 \hline
 B850 dipole moment (D)& 7.5 & 8.25 & 8.25 & 8.25   \\
 max. B850-B850 interaction (\cm) & 629 & 761 & 761 & 761   \\
$\hat{\Gamma}_{850}$ (\cm) & 0 & 0 & 1750 & 1750   \\
 \hline
 max. B850-B800 interaction (\cm) & 60 & 66 & 66 & 132   \\
  SB scaling $a_{800}=a_{850}$ & 0.3 & 0.3 & 0.02 & 0.02 \\
 \hline
\end{tabular}
\end{table*}

The absorption spectrum for {LH2} in a buffer/glycerol matrix also reported in Ref.~\cite{lohner15_23} looks rather different from the one obtained for the polymer matrix. In fact it is closer to the spectrum reported in Ref.~\cite{schroter18_1340}, also measured in glycerol, but at 77 K (see Fig.~\ref{fig:absorption}). In order to fit this spectrum and the dynamics reported in Ref.~\cite{schroter18_1340} we have designed three more models, which as far as the geometry is concerned build on model 1. In all three models we assume equal monomeric transition dipole moments (8.25~D) and introduce some heterogeneity by shifting the B800B monomeric site energies to 12650 \cm. 
This accounts for the different pigment-binding pockets in the two $\alpha$-apoproteins. Further, in \textbf{model 4} the B850-B800 couplings have been uniformly scaled by a factor of two (see Tab.~\ref{tab:lh2parameters}).
For all models inhomogeneous broadening is accounted for using the model of diagonal static disorder, which assumes an independent Gaussian distribution of site energies with variance of 150~\cm. The results presented below have been obtained by averaging over 5000 realizations.
 Other differences relate to the system-bath coupling, which is introduced in the following section.

\subsection{Exciton Dynamics}

The dynamics and spectroscopy of the  LH2 models will be treated using the standard system-bath (SB) model~\cite{may11,valkunas13,kuhn18_259}, see in particular the implementation in Ref.~\cite{kuhn97_4154}. The system part consists of the Frenkel exciton Hamiltonian  \textcolor{black}{(for the labeling see also Fig. S1 in the Supplementary Material)}
\begin{equation}
	H_{\rm S}=\sum_{mn}(\delta_{mn}E_m + J_{mn}) |m\rangle\langle n|
\end{equation}
with site energies, $E_m$, and Coulomb couplings, $J_{mn}$, as specified in the previous section. The single exciton eigenstates with energies, $E_\alpha$, will be expressed as
\begin{equation}
\label{eq:eigen}
\vert \alpha\rangle=\sum_m c_{m,\alpha}\vert m\rangle\,.
\end{equation}
The exciton dynamics is driven by interaction with an external laser field $\vec E (t)$ via the coupling to the transition dipole moments $\vec{\mu}_{m}$, i.e. 
\begin{equation}
H_{{\rm F}}(t)= - \sum_m 	\vec E (t)\vec{\mu}_{m} |m\rangle\langle 0| + {\rm h.c.}
\end{equation}
The transition dipole matrix elements in terms of the eigenstates are given by 
\begin{equation}\label{eq:dipole}
\vec{\mu}_{\alpha}=\sum_{m}\vec{\mu}_{m}c_{m,\alpha} \,.
\end{equation}
The exciton system is coupled to a thermal bath composed of harmonic oscillators with coordinates $q_\xi$ and frequencies $\omega_\xi$. The SB coupling is taken to be of the form
\begin{equation}
	H_{\rm SB}= \sum_m \sum_\xi \hbar \omega_\xi (g_{m, \xi}^{(1)} q_\xi + g_{m, \xi}^{(2)} q_\xi^2)|m\rangle\langle m| \, .
\end{equation}
\begin{figure}
\centering
	\includegraphics[width=\columnwidth]{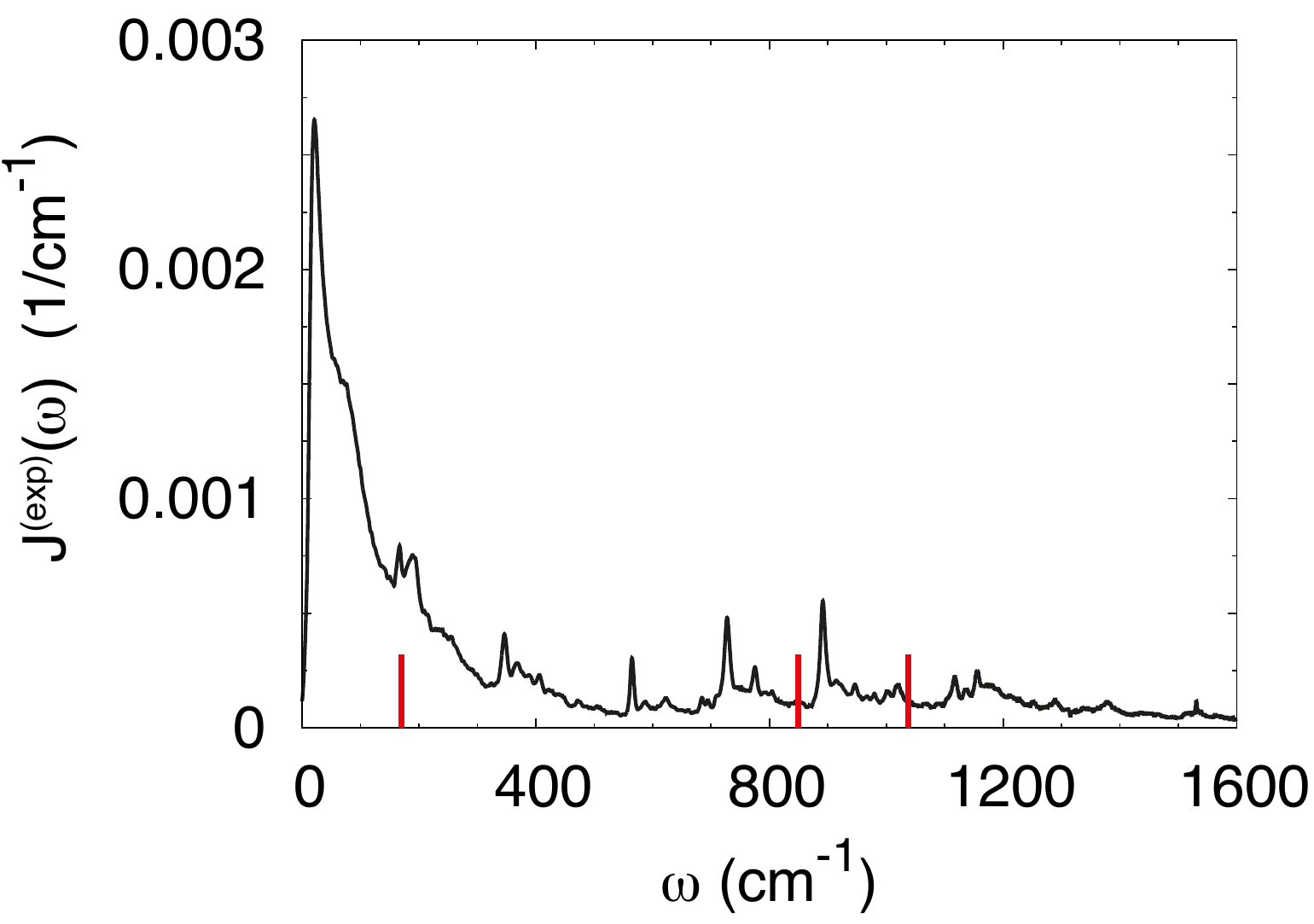}
	\caption{BChl $a$ spectral density used in this work. The data were obtained by fluorescence line-narrowing experiments in Ref.~\cite{ratsep11_024506}. The red vertical sticks are drawn at the energy gaps between the three peaks of the absorption spectrum of model 4 at 172, 864, and 1036 \cm.}
	\label{fig:SD}
\end{figure}

Here, the coupling strength for linear and quadratic coupling is $g_{m, \xi}^{(1)}$ and $g_{m, \xi}^{(2)}$, respectively.
\textcolor{black}{
The linear coupling constant is related to the Huang-Rhys factor via $S_{m, \xi}=(g_{m, \xi}^{(1)})^2/2$. Its effect is described by a spectral density $J_m(\omega) = \sum_\xi  S_{m,\xi} \delta(\omega-\omega_\xi)$. Central to phase and energy relation rate is the bath correlation function given by 
\begin{equation}
	C_m(\omega) = 2 \pi \omega^2[1+n(\omega)](J_m(\omega)-J_m(-\omega)]\, ,
\end{equation}
where $n(\omega)$ is the Bose-Einstein distribution function. In the following the shape of the 
spectral density is taken taken from the fluorescence line-narrowing experiment on BChl $a$ in solution~\cite{ratsep11_024506} called $J^{\rm (exp)}(\omega)$, see Fig.~\ref{fig:SD}. Its reorganization energy amounts to $\lambda^{\rm (exp)}=\hbar \int d\omega \omega J^{\rm (exp)}(\omega) =196$~\cm. This choice provides a spectral density having a form which is typical for intramolecular modes of  BChl $a$, but it doesn't capture  effects due to the specific environment in LH2. In passing we note that the specific values of the spectral density at the relevant transition frequencies (see sticks in Fig.~\ref{fig:SD}) determine the relaxation rates. This means that using a simple Debye fit while retaining the overall reorganization energy would change the ratio of the spectral density values if comparing different regions and thus the pattern of relaxation rates. }  

\textcolor{black}{In the simulations reported below,
to have some flexibility concerning the fitting of relaxation and dephasing rates we will introduce a site dependent  SB scaling parameter $a_m$, i.e. the spectral density is assumed to have the form $J_m(\omega)=a_m  J^{\rm (exp)}(\omega)$. In fact, the best agreement with experiment has been obtained using the same scaling parameter for all sites. Note that in view of the simplicity of the present exciton-vibrational model this empirical parameter should not be overinterpreted. In fact, of more relevance are the actual relaxation rates which also contain the effect of excitonic couplings, see below. }
As far as the quadratic coupling is concerned we will restrict ourselves to the pure dephasing contribution only (see also Ref.~\cite{kuhn97_4154}). 

The dynamics of the reduced exciton density operator, $\rho$,  will be treated using the Redfield model in Bloch approximation~\cite{may11}
\begin{equation}\label{eq:qmewithfield}
\dfrac{d}{dt}\rho(t)=-\frac{i}{\hbar}\left[H_{\rm{S}}+H_{\rm{F}}(t),\rho(t)\right]-R\rho(t)
\end{equation}
with the relaxation matrix $R$ having contributions for population relaxation
\begin{equation}
	R_{\alpha\alpha,\beta\beta}=-k_{\beta\rightarrow\alpha}+\delta_{ \alpha\beta}\sum_{\gamma}k_{\alpha \rightarrow\gamma}
\end{equation}
and coherence dephasing
\begin{equation}
	R_{\alpha\beta,\alpha\beta}=\hat{\Gamma}_{\alpha\beta}+\dfrac{1}{2}\sum_{ \gamma\neq\alpha}k_{\alpha\rightarrow\gamma}+\dfrac{1}{2}\sum_{ \gamma\neq\beta}k_{\beta\rightarrow\gamma} \,.
\end{equation}
Here, the energy relaxation rates between states $\alpha$ and $\beta$ are given by 
\begin{equation}\label{eq:dampingab}
k_{\alpha\rightarrow \beta}=\sum_{m}C_{m}(\omega_{\alpha\beta})\vert c_{m,\alpha}\vert^2\vert c_{m,\beta}\vert^2
\end{equation}
and the pure depasing rates are
\begin{align}
&\hat{\Gamma}_{\alpha \beta}=\sum_{m}\hat{\Gamma}_m (\vert c_{m,\alpha}\vert^2-\vert c_{m,\beta}\vert^2)^2 \, \\
&\hat{\Gamma}_{\alpha 0}=\sum_{m}\hat{\Gamma}_m\vert c_{m,\alpha}\vert^4 \, ,
\label{eq:pure}
\end{align}
with $\hat{\Gamma}_m \propto |g_{m,\xi}^{(2)}|^2$ being the pure dephasing rate for molecule $m$, which is treated as a parameter.
Thus, the phase relaxation rates for the excitonic transitions from the ground state read
\begin{equation}\label{eq:oerelaxrate}
\gamma_{\alpha}=\sum_{\beta\neq\alpha}k_{ \alpha\rightarrow\beta}+2\hat{\Gamma}_{\alpha 0} \,. 
\end{equation}

\section{Results}
\label{sec:results}
\textcolor{black}{
In the following we will present results obtained for four models; cf. Tab.~\ref{tab:lh2parameters}. Starting from a model similar to the one in Ref.~\cite{lohner15_23} (\textbf{model 1}) supplemented by energy and phase relaxation, by changing the model parameters we will eventually arrive at \textbf{model 4}, which reproduces both the absorption spectrum and the energy relaxation rates obtained in the experiment~\cite{schroter18_114107}. The other two models (2 and 3) merely serve to illustrate the effect of the different parameters on the absorption spectrum.}
\subsection{Absorption Spectra}
\begin{figure}[t]
\begin{center}
\includegraphics[width=1\columnwidth]{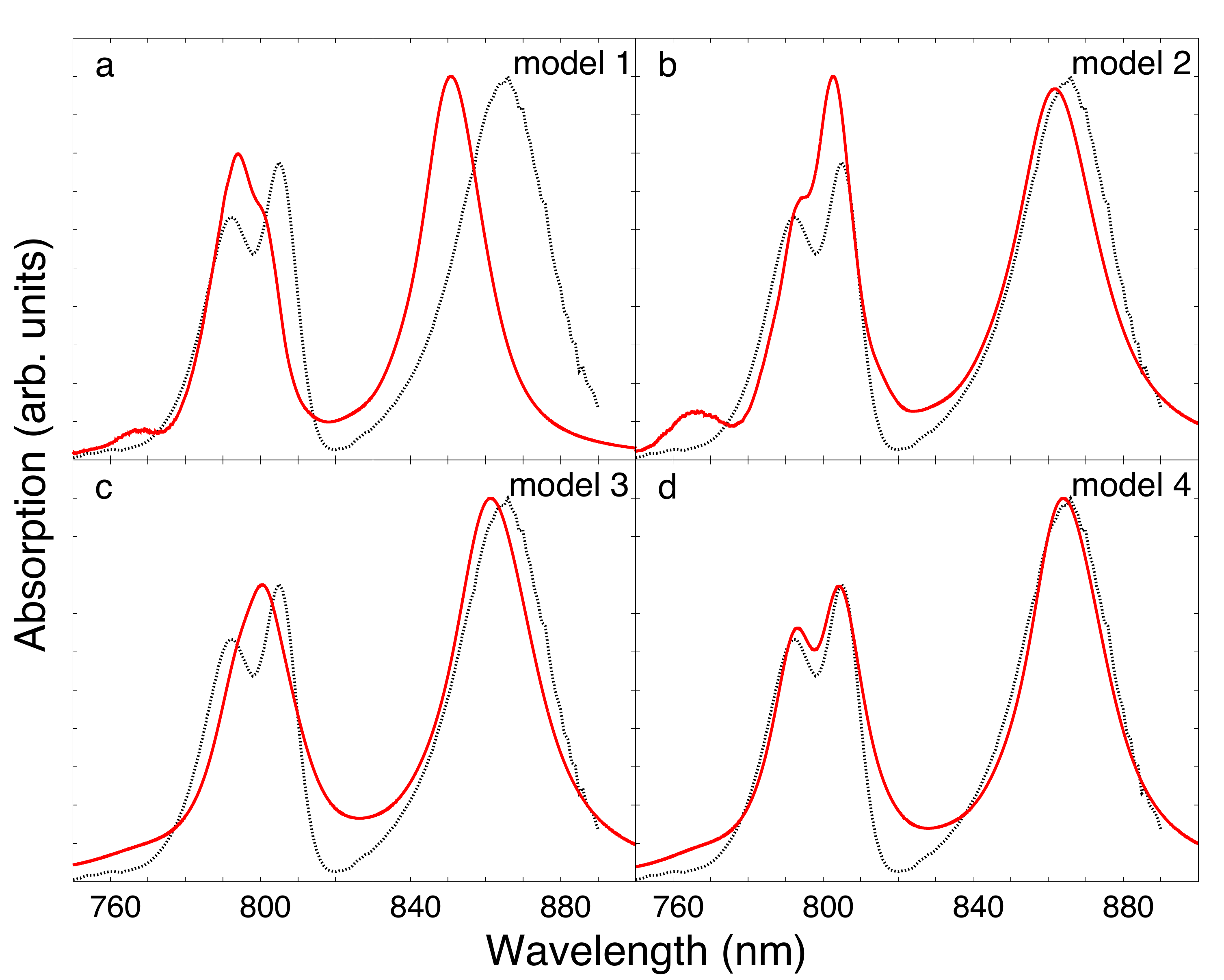}
\caption{ Absorption spectra (red full line) as obtained for the different models upon averaging over 5000 samples of an uncorrelated  Gaussian distribution of local transition energies.  Also shown is the experimental spectrum (black dashed line) from Ref.~\cite{schroter18_1340}. Model parameters are given in Tab.~\ref{tab:lh2parameters}.}
\label{fig:absorption}
\end{center}
\end{figure}

In the following we will discuss the absorption spectrum ($T=77$~K)
\begin{equation}\label{eq:absorptionlh2}
A(\omega)=\left\langle\sum_{\alpha}\dfrac{ \gamma_{\alpha}\vert \vec{\mu}_{\alpha}\vert^2}{(\omega-\omega_{\alpha})^2+\gamma^2_{\alpha}/4} \right\rangle_{\rm{disorder}}
\end{equation}
to fully specify the four different models according to Tab.~\ref{tab:lh2parameters}. In Fig.~\ref{fig:absorption}a the experimental absorption spectrum~\cite{schroter18_1340} is compared with a simulation using the original model of Ref.~\cite{lohner15_23}, supplemented by the excitonic phase relaxation. Here, we did not include pure depashing and tuned the SB scaling parameters $a_m$ such as to give a reasonable fit to the experimental linewidths (\textbf{model 1}). It turns out that model 1, which was paramaterized in Ref.~\cite{lohner15_23} to reproduce the 1.2~K matrix spectra, gives only a poor agreement with the 77~K glycerol spectra. First, the B800-B850 splitting is too small and, second, the B800 band splitting has a reversed order of intensities.

To improve the agreement, first, it was found that the dipole moment for the B850  {BChl} $a$ should be increased from 7.5 D to 8.25 D (the same as B800) to match the B800-B850 splitting. Further, once an energy shift $\Delta E_{\rm 800B}=-250$~\cm{} is introduced the ratio of the B800 peak heights is reversed, which is shown in Fig.~\ref{fig:absorption}b. In fact, introducing an energy shift between B800 and B850 subunits will also reproduce the splitting between B800 and B850, however, the ratio of the B800 peaks cannot be matched due to the interdependence between this ratio and the energy shift. This set of parameters defines \textbf{model 2}; see Tab.~\ref{tab:lh2parameters}.

Comparing experimental and calculated absorption spectra, one notices that the ratio of the B800 to B850 peak heights does not match. In addition there is an  extra peak in the calculation near 765 nm. The intensity ratios can be influenced by the linewidths, i.e. the SB coupling parameters. Including   pure dephasing, the 765~nm peak can be suppressed and the widths of the B800 and B850 peaks can be reasonably matched (see Fig.~\ref{fig:absorption}c) with the parameters of \textbf{model 3} as given in Tab.~\ref{tab:lh2parameters}. 

Inspecting Fig.~\ref{fig:absorption}c we notice that the B800 peak splitting is still not reproduced. In principle, two factors have a direct influence on this peak splitting, which are the energy shift $\Delta E_{\rm 800B}$ and the inter-pool coupling, $J_{800-850}$, between B800 and B850 monomers. Increasing  $\Delta E_{\rm 800B}$  will also cause a decrease of the B800-B850 gap. However, it is found that a scaling of all  couplings of type $J_{800-850}$ by a factor of two in \textbf{model 4} gives the best agreement with the experimental absorption spectrum as shown in Fig.~\ref{fig:absorption}d. Note that the actual values for $a_m$ and $\hat \Gamma_m$ have been fixed using the population dynamics (see below),   i.e. while the spectrum is influenced by both parameters, the population flow depends on $a_m$ only.

\begin{figure}[t]
\begin{center}
\includegraphics[width=1\columnwidth]{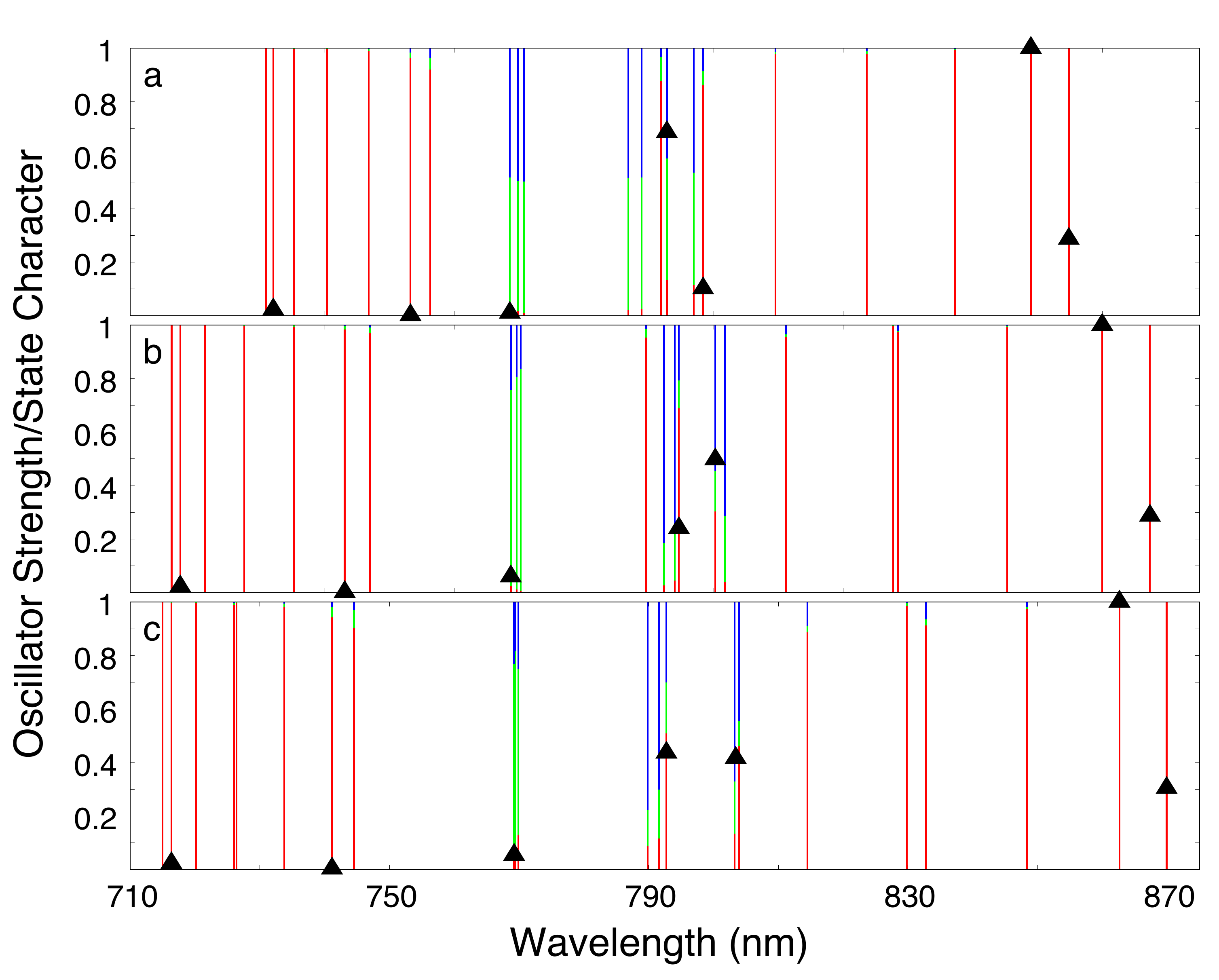}
\caption{Eigenvalues, state character as measured by the coefficient $c_{\alpha}(i)=\sum_{m\in i}\vert c_{m,\alpha}\vert^2$ (red: $i=$B850, green: $i=$B800A, blue: $i=$B800B), and oscillator strength (black triangles) for model 1 (a), models 2,3 (b), and model 4 (c). The calculations have been performed without disorder.} 
\label{fig:lh2eigen}
\end{center}
\end{figure}

 In order to unravel the changes in the spectra for the different models, the eigenstates, Eq.~\eqref{eq:eigen}, and oscillator strengths $|\mu_\alpha|^2$, Eq. \eqref{eq:dipole}, will be analyzed for the case of no disorder. Here, a decomposition of the eigenstates in terms of local B800 and B850 state has been performed according to $c_{\alpha}(i)=\sum_{m\in i}\vert c_{m,\alpha}\vert^2$ with $i=$(B800A, B800B, B850). The results are shown in  Fig.~\ref{fig:lh2eigen}. For all models the total width of the  eigenstate spectrum is determined by B850-like states. Further, due to the relatively strong coupling between the B800 monomers, the band structure related to these monomers is clearly discernible. The B800 band is approximately located at the overall band center. 
 Due to the high symmetry, oscillator strengths is distributed over a few transitions only, e.g. notably at the lower band edge (B850-like states). As far as the B800 double peak is concerned, the mixing between B800- and B850-like states is of prime importance. In the original model 1, the transition at the blue side of the double peak (B800b) is dominantly of B800 origin, while the red peak (B800r) is of B850 origin. Going to models 2 and 3, where the local transition dipoles are equal and where there is a shift $\Delta E_{\rm 800B}$ of the B800B monomer energies, reverses this assignment along with a reversed transition strength ratio. Increasing the B800-B850 coupling, increases the splitting between the bright states. At the same time they become more B800-like. Here, B800b has about equal contributions from B800A and B800B, whereas B800r is dominated by B800B excitations.

 \begin{figure}[t]
\begin{center}
\includegraphics[width=\columnwidth]{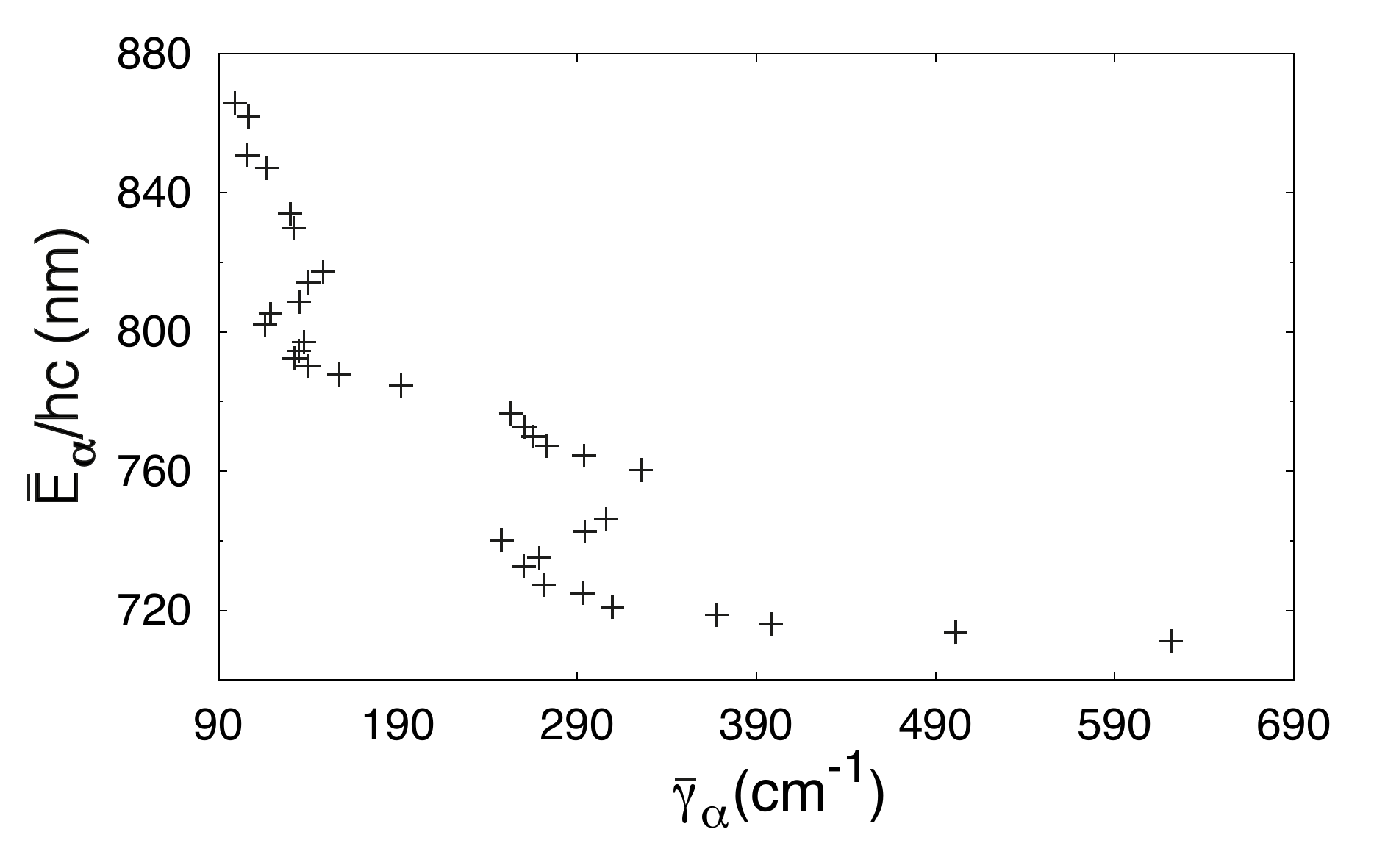} 
\caption{Phase relaxation rates $\bar \gamma_{\alpha}$ and eigenenergies $\bar E_{\alpha}$ after averaging over an inhomogeneous ensemble (5000 realizations) for model 4.}
\label{fig:lh2gam}
\end{center}
\end{figure}

The effect of the linewidth on shaping the overall spectrum can be appreciated by inspecting Fig.~\ref{fig:lh2gam}. It shows the phase relaxation rates averaged with respect to the Gaussian distribution of site energies, $ \bar \gamma_{\alpha}$, in dependence on the average energies $\bar E_{\alpha}$ for model 4. Similar to the model discussed for \textit{Rps. acidophila} in Ref.~\cite{kuhn02_15} the relaxation rate increase with increasing energy, being largest at the upper edge of the exciton band. This fact is responsible for suppression of the peak near 765~nm. The reason for this behavior is that  with increasing energy the number of relaxation channels increases as well. The non-monotonous behavior is due to the change in delocalization when moving through B800 and B850 dominated bands. Overall we notice that the rather high values for the local pure dephasing constants should be taken with caution due to the influence of the eigenstate coefficients in the final dephasing rates, Eq.~\eqref{eq:pure}.

 \begin{figure}[t]
\begin{center}
\includegraphics[width=0.9\columnwidth]{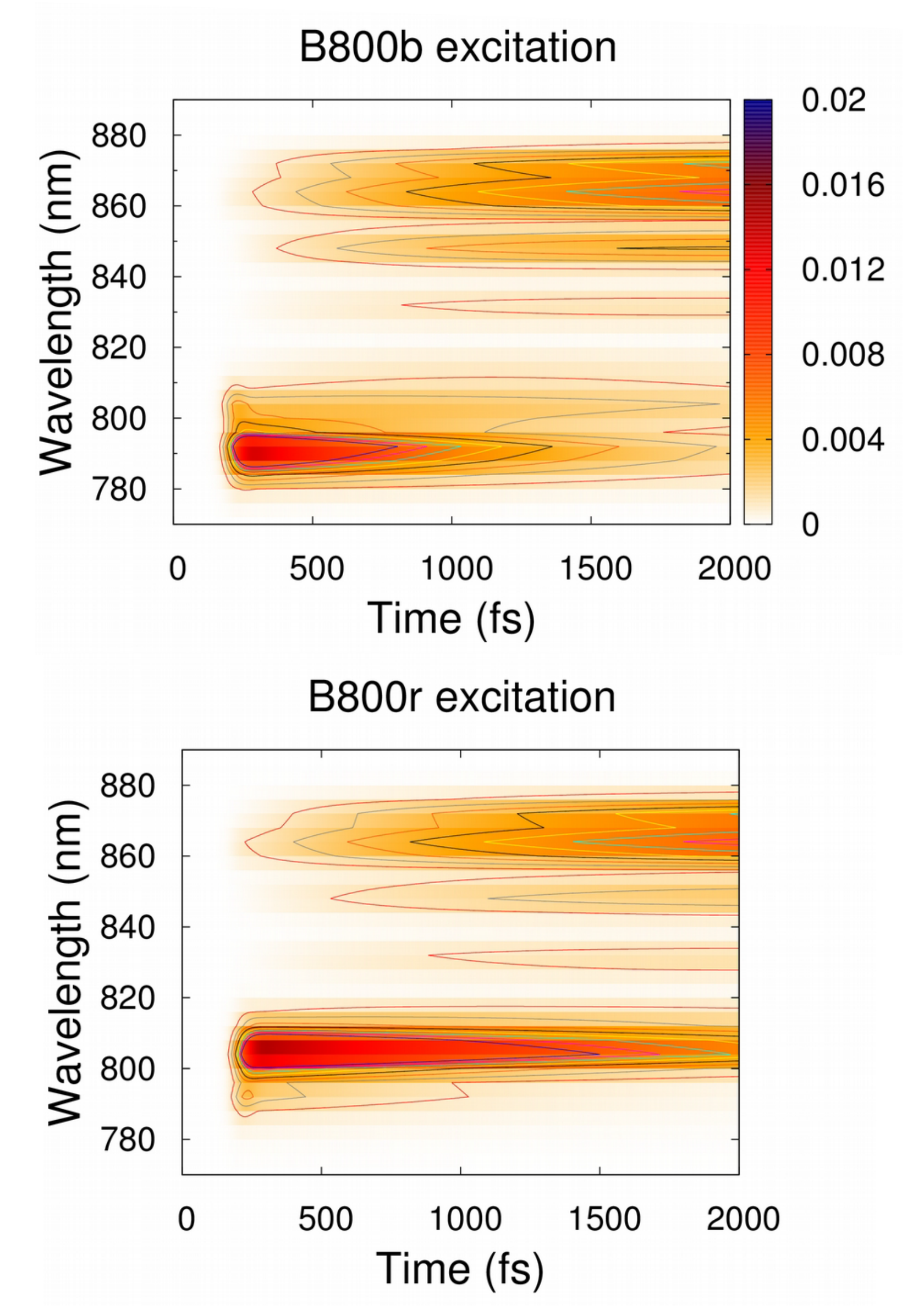}
\caption{The populations dynamics of model 4 in the range from  760 nm to 900 nm (step 4 nm) for excitation at the two B800 band maxima (contour values from 0.001 to 0.01 by 0.001). \textcolor{black}{For a plot of the populations scaled by the transition dipole strength of the respective states, see Fig. S2 in the Supplementary Material}} 
\label{fig:pop2D}
\end{center}
\end{figure}
\subsection{Population Dynamics} 
In the following we will investigate whether the parametrization of model 4 is in accord with the time scales of energy relaxation obtained in Ref.~\cite{schroter18_1340}. To this end, the exciton dynamics driven by a Gaussian-shaped laser pulse is studied, i.e.
\begin{equation}\label{eq:lh2extfield}
\vec{E}(t)=\vec{e} E_{0}\cos(\omega t)\exp\left(- \dfrac{(t-t_0)^2}{2\sigma^2}\right)\, .
\end{equation}
 Here, a field strength of ${E}_{0}=$1.1$\times 10^7$ V/m is chosen such as to give about 5\% excited state population for $t_0=200$ fs and $\sigma=42.5$ fs (i.e. the FWHM of the pulse is 100 fs).

\begin{figure}[t]
\begin{center}
\includegraphics[width=0.9\columnwidth]{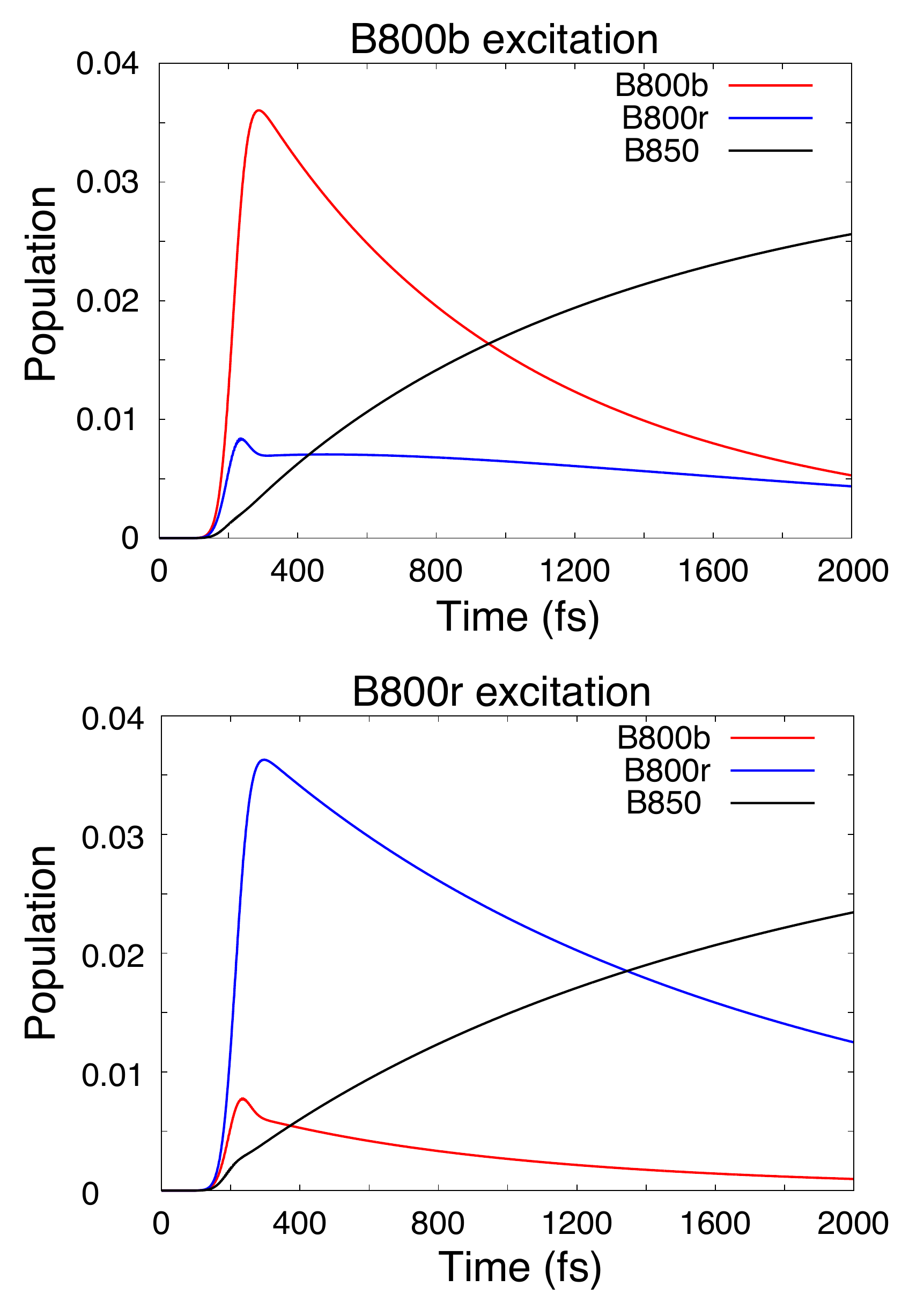}
\caption{The integrated population for the three peaks in the  {LH2} absorption spectrum, i.e. B800b:[788,800] nm, B800r:[800,812] nm and B850:[852,876] nm and for the two excitation conditions of Fig.~\ref{fig:pop2D}.} 
\label{fig:pop1D}
\end{center}
\end{figure}

To account for the averaging over disorder, population dynamics will be assigned to certain wavelength ranges as $P_{ab}=\sum_{\alpha}\rho_{\alpha\alpha}$ if $E_{\alpha}/hc \in [\lambda_a,\lambda_b]$ for a given sample. To focus on the time scales associated with the excitation of the two peaks of the  B800 band, two excitation cases are introduced as follows: the case B800b/B800r excitation corresponds to excitation within the wavelength range [788,800]~nm/ [800,812]~nm. In other words, B800b and B800r matches the lower and higher wavelength peak, respectively. In both cases  the direction, $\vec e$, and frequency, $\omega$, of the external field are assumed to be the same as the direction and eigenvalue, respectively,  for the largest dipole moment $\vec{\mu}_{\alpha}$ in the considered frequency range. The population dynamics will be analyzed in terms of contour plots, Fig.~\ref{fig:pop2D}, and integrated frequency intervals, Fig.~\ref{fig:pop1D}.

The population dynamics for the two cases during 2000 fs is shown in Fig.~\ref{fig:pop2D}. First, let's consider excitation of the lower wavelength band (B800b), cf. upper panel of Fig.~\ref{fig:pop2D}. Here, the states in B800b are dominantly excited and in B800r are weakly excited by the external pulse. After the pulse, the excitation energy quickly transfers from B800b to B850. However, it is found that there is no apparent reduction of the populations in the B800r range from 400 fs to 800 fs, and only after 1200 fs appreciable depopulation sets in. This   is more clearly observed from the integrated populations in the upper panel of Fig.~\ref{fig:pop1D}. The reason is that shortly after the pulse the direct relaxation from B800b to B800r keeps the populations in the B800r range approximately unchanged, but after some time there is not enough population flow to the B800r range anymore to compensate for the transition from B800r to B850. 

Next, we focus on the case where the higher wavelength band is excited (B800r), cf. lower panels of Figs.~\ref{fig:pop2D} and \ref{fig:pop1D}. Here, the states in B800b are only weakly excited, while those in B800r are strongly excited by the external pulse. There is no plateau-like behavior for the B800r population and both bands decay with different time scales.

The depopulation times of the B800r and B800b band states can be obtained from the integrated populations  in Fig.~\ref{fig:pop1D}. To this end,  a fit of the populations to a kinetic three state model has been performed. This gives a time scale of 1.13~ps for the direct B800b to B850 relaxation as  well as 3.25~ps and 1.43~ps for the two step relaxation B800b$\rightarrow$B800r$\rightarrow$B850.
 %
\begin{figure}[t]
\begin{center}
\includegraphics[width=0.95\columnwidth]{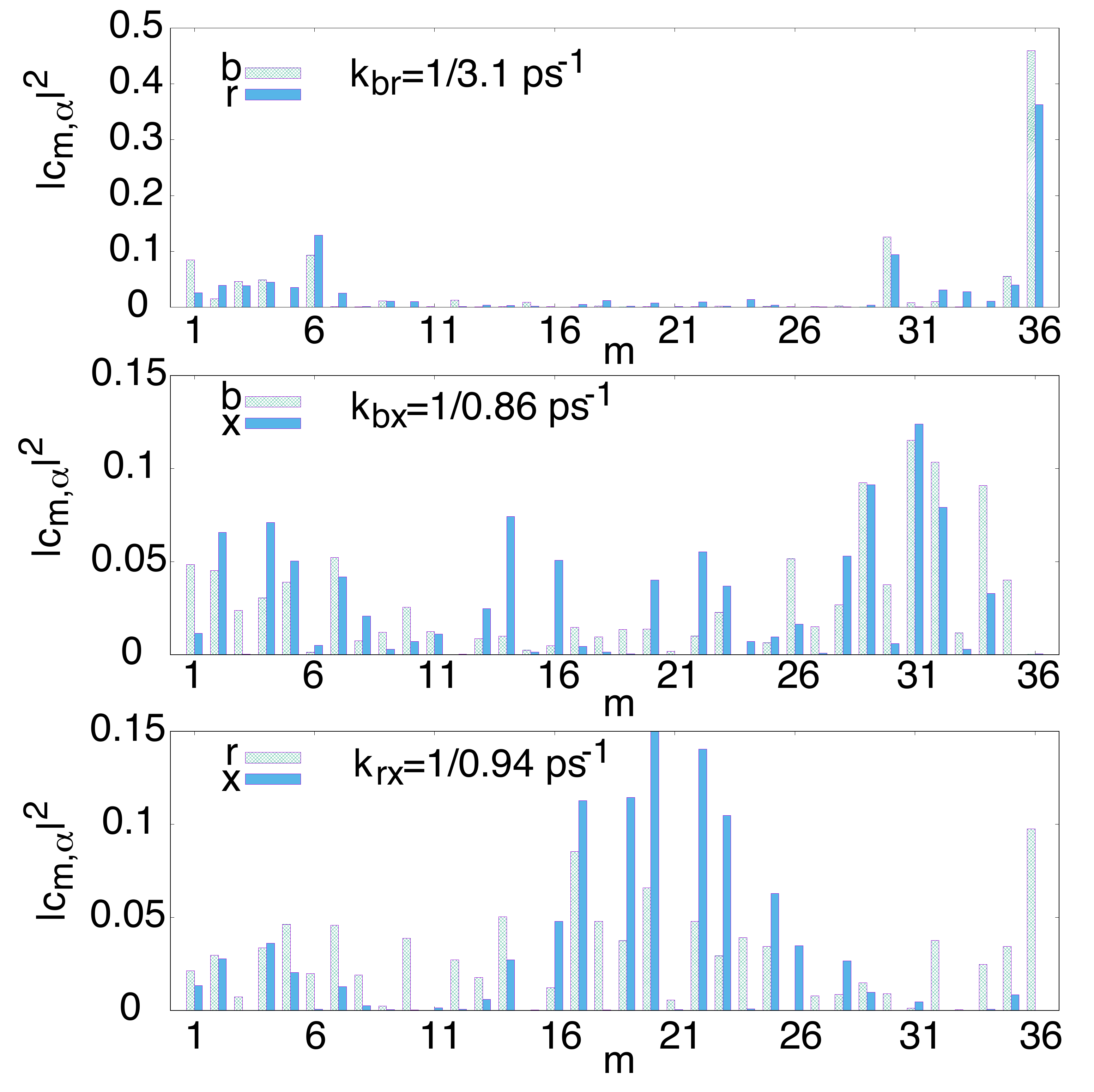}
\caption{Analysis of the relaxation rates, Eq.~\eqref{eq:dampingab}, in terms of the eigenfunction coefficients for particular (different) disorder realizations, chosen such as to resemble the values obtained   for transitions between the B800b and B800r band (upper panel), between the B800b and B850 bands (middle panel), and between the B800r and the B850 band (lower panel). The accepting B850 states are labeled by $x$. For $k_{bx}$ and $k_{rx}$ initial/final states are at 789/849 nm  and 804/861 nm, respectively. For $k_{br}$ the values are 792~nm and 805 nm. The labeling follows the sequence B850, B850, B800A, B850, B850, B800B etc. \textcolor{black}{Note that the site energies of the particular disorder realizations used here are given in Fig. S3 of the Supplementary  Material.)}
} 
\label{fig:rates}
\end{center}
\end{figure}

The nature of these relaxation processes can be unraveled by analyzing the relaxation rates, Eq.~\eqref{eq:dampingab}. This has been done in Fig.~\ref{fig:rates} for particular disorder realizations, which have been chosen such as to resemble the values of the decay rates for the ensemble. Of course, analysis of a single member of the ensemble should  not be over-interpreted and at best provides a qualitative picture. The relaxation rates depend on the spectral density taken at the transition frequency as well as on the wave function overlap $|\langle\alpha|m\rangle\langle m | \beta \rangle|^2$. According to Fig.~\ref{fig:SD} the spectral density changes by a factor $\sim$5  when comparing its values at the transition frequencies between the B850 and the B800  bands. The eigenfunction coefficients are shown in Fig.~\ref{fig:rates}. Overall, we notice that comparing the B800b to B800r relaxation with the decay of B800b/B800r towards the B850 states, in the latter cases there are substantially more local states involved (i.e.\ the eigenstates are more delocalized), which  overcompensates the smaller value of the spectral density.

For the B800b to B800r relaxation we find large overlaps at sites 6, 30, 36, i.e. this relaxation is dominated by B800B local states. This is in accord with the fact that both absorption peaks of the B800 band are to a good extent of B800B origin (cf. Fig.~\ref{fig:lh2eigen}).
The decay of the B800b band states is due to the mixing between local B800A/B and B850 transitions. In some case there is only an amplitude at the local B800A/B site (3, 18, 27, 30), but no such amplitude in the final state. The main contribution to the relaxation rate comes from state pairs which both have local B850 amplitudes. For this particular disorder realization, the exciton eigenfunctions responsible for relaxation are delocalized on the segment  with $m=29, 31, 32$. The initial state for the relaxation of the B800r band states has local amplitudes on the B800B and B850 sites, and little involvement of the higher energetic B800A site. Responsible for the decay is a pair of states delocalized on the segment with $m=17, 19, 20, 22, 23$.

\begin{figure}[t]
\begin{center}
\includegraphics[width=0.65\columnwidth]{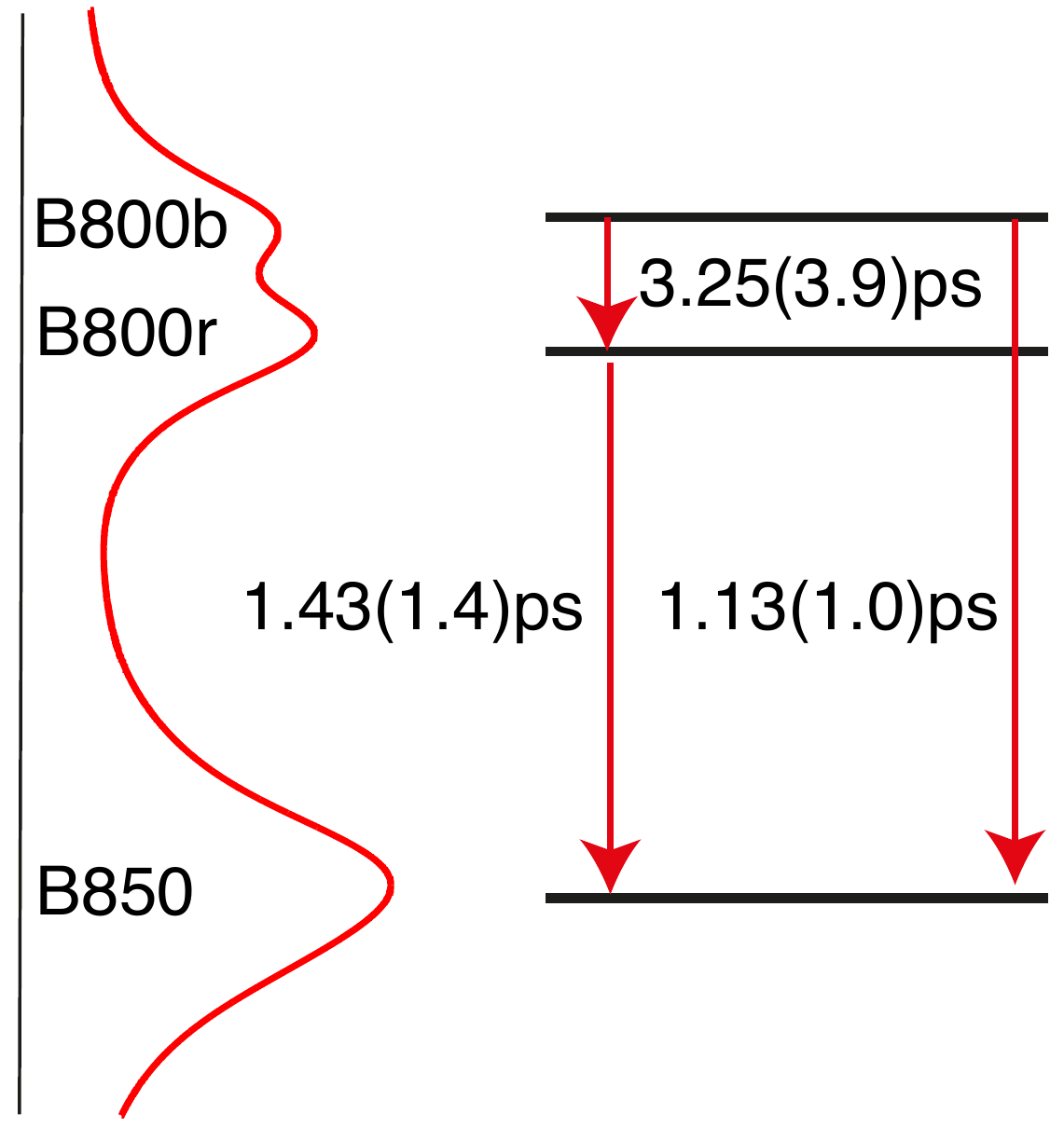}
\caption{Calculated absorption spectrum (left) and relaxation times for model 4 assuming an effective 3-level scheme (in parenthesis experimental results from Ref.~\cite{schroter18_1340}). } 
\label{fig:scheme}
\end{center}
\end{figure}

\section{Summary}
\label{sec:summary}
The peculiar double-peak structure of the B800 band of \textit{Alc.\ vinosum} at 77~K has been investigated from the perspectives of absorption spectroscopy and exciton population dynamics. Thereby, it has been assumed that the key structural feature is a dimerization of the B800 pool, in accord with previous experimental and theoretical studies~\cite{lohner15_23,schroter18_1340}. The structural model, previously developed by L\"ohner et al.~\cite{lohner15_23} for 1.2~K polymer matrix conditions,  has been adapted and extended to include dephasing and energy relaxation within a system-bath model. Simulations of absorption spectra and population dynamics have been performed for different models using Redfield relaxation theory.

A parametrization of a structural and system-bath model has been identified, which reproduces the spectrum as well as the population dynamics in good agreement with experiment. The relevant results are compiled in Fig.~\ref{fig:scheme}. On the one hand side, this can be viewed as advanced fitting of multiple sets of experimental data. On the other hand side, the analysis of the results provided inside into the details, which could be operative for this particular LH2 complex.  In fact, the LH2 of \textit{Alc. vinosum} features an interesting interplay of two excitonic bands, which are originating from different pigment pools. This involves state, which for the uncoupled pools are essentially optically dark by symmetry. The particular double-peak structure of the B800 band is emerging due to B800 dimerization but also due the coupling to the B850 pool, causing a particular state mixing and thus a sensitivity to resonances and couplings strengths between the exciton manifolds of the separate pools. In terms of the population dynamics, this opens different relaxation channels upon excitation of the B800 band. In particular excitation of the short wavelength peak (B800b) leads to relaxation to the B850 band via two pathways: direct transfer to B850 with a time scale of $\sim$1.1~ps, which is the main pathway, and indirect slower two-step transfer via B800r to B850. 

\textcolor{black}{The present model has to be viewed as a suggestion based on advanced fitting. The crucial point for obtaining the good agreement with experiment has been the \textit{ad hoc} change of the Coulomb coupling strength between the B800 and B850 pools. In terms of the structure proposed in Ref. \cite{lohner15_23} this would correspond to a decrease of the vertical distance between the B800 and B850 rings from 17.3 \AA{} to 14 \AA. Other factors which could influence the Coulomb coupling are the different screening of the inter- and intrapool interactions as well as the breakdown of the dipole approximation. Without further structural information it will be difficult to disentangle these effects in their influence on the simulation results.}
\section{Acknowledgments}
The authors thank the Deutsche Forschungsgemeinschaft (DFG) for financial support through the Sfb 652.

%

%
\end{document}